\documentclass{ws-procs10x7}
\usepackage{balance}

\newcolumntype{d}[1]{D{.}{.}{#1}}

\def\Journal#1#2#3#4{{\it #1} {\bf #2}, #3 (#4)}

\makeindex
\begin{document}

\title{ANALYSIS OF THE GENERATION OF PHOTON PAIRS\\
IN PERIODICALLY POLED LITHIUM NIOBATE}

\author{J. S\"{o}derholm, K. Hirano, S. Mori \and S. Inoue$^*$}

\address{Institute of Quantum Science, Nihon University,\\
1-8 Kanda-Surugadai, Chiyoda-ku, Tokyo 101-8308, Japan\\
$^*$E-mail: ino@phys.cst.nihon-u.ac.jp}

\author{S. Kurimura}

\address{National Institute for Materials Science, 1-1 Namiki, Tsukuba-shi,
Ibaraki 305-0044, Japan}

\twocolumn[\maketitle\abstract{The process of spontaneous parametric
down-conversion (SPDC) in nonlinear crystals makes it fairly easy to
generate entangled photon states. It has been known for some time
that the conversion efficiency can be improved by employing
quasi-phase-matching in periodically poled crystals. Using two
single-photon detectors, we have analyzed the photon pairs generated
by SPDC in a periodically poled lithium niobate crystal pumped by a
femtosecond laser. Several parameters could be varied in our setup,
allowing us to obtain data in close agreement with both thermal and
Poissonian photon-pair distributions.} 
]

\section{Introduction}

Through optical nonlinearities in a crystal, one pump photon can
be converted into two photons with longer wavelengths. The fact
that these down-converted photons are always generated in pairs
has made it possible to use them for many fundamental
quantum-mechanical experiments.\cite{RMP} These photon pairs are
also suitable for fundamentally secure quantum-key distribution,
which is usually referred to as quantum cryptography.\cite{Gisin}
Recently, it was recognized that the efficiency of the photon-pair
generation can be improved by using periodically poled
crystals.\cite{PPLN} The ferroelectric polarization of these
crystals is periodically inverted, resulting in flips of the sign
of the nonlinear coupling $\chi^{(2)}$. Through
quasi-phase-matching one may then exploit larger nonlinear
coefficients than what is possible using birefringent phase
matching.

In the present work, we have used single-photon detectors to
analyze the down-converted photon pairs generated in a
periodically poled lithium niobate (PPLN) crystal under various
conditions.

\section{Experimental setup}

\begin{figure}[b]
\centerline{\psfig{file=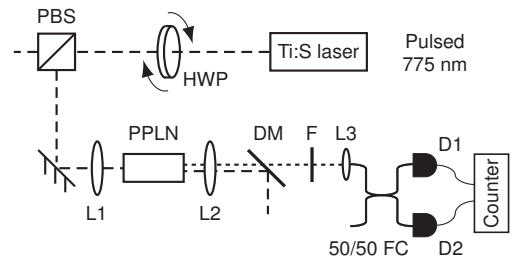,width=2.6in}} \caption{Schematic
drawing of our experimental setup used to analyze the photon pairs
generated in a PPLN crystal. (See text for explanation.)}
\label{fig:Setup}
\end{figure}

The experimental setup is schematically depicted in
Fig.~\ref{fig:Setup}. Our 10-mm-long bulk PPLN crystal was doped
with 5~mol\% MgO and pumped by 200-fs-long pulses from a
Titanium:sapphire laser with a repetition rate of 80 MHz. The
poling periodicity of the crystal was 19.4~$\mu$m, corresponding
to degenerate quasi-phase-matched parametric down-conversion at
1544~nm. However, in order to make the frequency of the degenerate
photon pairs coincide with our interference filters' central
frequency of 1550~nm, the wavelength of the laser was set to
775~nm. A half-wave plate (HWP) and a polarizing beam splitter
(PBS) were used to attenuate the laser power without distorting
the laser mode and pulse. As the polarization direction of the
laser beam depended on the rotation angle of the HWP, the
intensity of the pump beam emerging from the PBS could be varied.

Three different pairs of lenses were used to focus the pump beam
in the crystal, and collect the down-converted light. In the case
with closest focus, we did not have two matching lenses at hand,
so the focal length of L1 was 8.2~mm, whereas that of L2 was
10~mm. In the other two cases, both L1 and L2 had a focal length
of 14.8~mm and 50~mm, respectively. Effective down-conversion was
restricted to the focus, since the pump intensity was much higher
there.

A dichroic mirror (DM) and a filter (F), with a
full-width-at-half-maximum bandwidth of either 10 or 30~nm, were
used to discriminate the down-converted light from the pump. The
down-converted light was subsequently coupled into a fiber using a
third lens (L3). After splitting the beam with a 50/50 fiber
coupler (50/50 FC), both single and coincidence events at the
single-photon detectors (D1 and D2) were recorded. The detectors
were gated at a rate of 316~kHz and synchronized with the pump
pulses.

\section{Theory}

Since our detectors cannot distinguish if one or more photons make
them ``click", the single-count rate at detector $k$ (when
neglecting dark counts) can be expressed as
\begin{equation}
S_k = R \sum_{m=0}^\infty \frac{p (m)}{4^{m}} \sum_{n=0}^{2 m}
\left( \begin{array}{c} 2 m \\ n \end{array} \right) [1 - (1 - T
\eta_k)^n] , \label{eq:Sk}
\end{equation}
where $R$ is the gate rate, $p (m)$ denotes the probability for
$m$ photon pairs to be generated, $T$ is the transmittivity of the
optical components, and $\eta_k$ is the detector efficiency.
Similarly, the coincidence-count rate, i.e., the rate with which
both the detectors click during the same gate pulse, can be
written as
\begin{eqnarray}
C & = & R \sum_{m=0}^\infty \frac{p (m)}{4^{m}} \sum_{n=0}^{2 m}
\left( \begin{array}{c} 2 m \\ n \end{array} \right) [1 - (1 - T
\eta_1)^n] \nonumber \\
& & \times [1 - (1 - T \eta_2)^{2 m - n}] . \label{eq:C}
\end{eqnarray}

\subsection{Photon-pair distributions}

In the present setup, most photon pairs will be composed of
photons with different wavelengths. With no initial photons in the
down-conversion modes, the nondegenerate SPDC process is known to
generate the two-mode squeezed vacuum state. The photon-pair
distribution is then thermal\cite{Mollow}
\begin{equation}
p_{\rm th} (m) = \frac{\mu^m}{(\mu + 1)^{m+1}} ,
\label{eq:ThermDistr}
\end{equation}
where $\mu = \sinh^2 r$ is the average number of photon pairs, and
the squeezing parameter $r$ is proportional to the electric field
of the pump.\cite{Scully} Assuming that the form of the pump pulse
is independent of the pump power, we thus obtain $\mu = \sinh^2
\sqrt{K P_{\rm ave}}$, where $K$ is a constant and $P_{\rm ave}$
is the average pump power. We note that $\mu \approx K P_{\rm
ave}$ as $\sqrt{K P_{\rm ave}} \ll 1$.

On the other hand, when the detected photons originate from many
distinguishable down-conversion processes, the photon-pair
distribution can be approximated\cite{deRiedmatten} by the
Poissonian distribution
\begin{equation}
p_{\rm poi} (m) = \frac{\nu^m e^{-\nu}}{m!} , \label{eq:PoiDistr}
\end{equation}
where $\nu$ is the average value of the total number of photon
pairs generated by a single pump pulse. As there are many
distinguishable processes, the average photon-pair number in each
of them is usually small ($\mu \ll 1$), and therefore proportional
to the average pump power. We then get $\nu \approx \mathcal{K}
P_{\rm ave}$, where $\mathcal{K}$ is a constant. In a recent
experiment\cite{Mori} similar to the one reported here, but
employing a semiconductor laser and a 3-cm-long PPLN waveguide, we
have been able to fit both single and coincidence counts versus
pump power to such a Poissonian photon-pair distribution.

\begin{figure}[t]
\centerline{\psfig{file=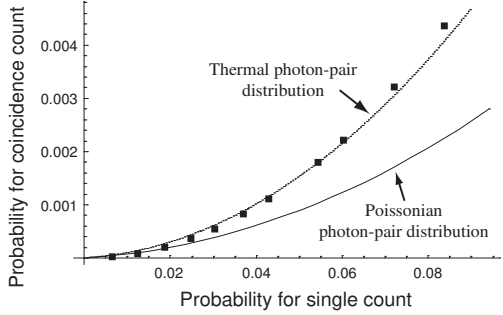,width=2.6in}}
\caption{Experimental results ($\blacksquare$) obtained with a
focus length of 8.2~mm and a 10-nm filter.} \label{fig:8mm}
\end{figure}

\subsection{Dark counts}

As there is a probability for the detectors to click even when no
photon is incident, the effects of these dark counts should be
compensated for before comparing the experimental data with the
expressions in the previous section. Noting that the raw
single-count rates $S_k^{(\mathrm{raw})}$ are due to either
incident photons, dark counts, or both, one finds that the
photon-induced single-count rates are given by
\begin{equation}
S_k^{(\mathrm{ph})} = \frac{S_k^{(\mathrm{raw})} - \delta_k}{1 -
\delta_k R^{-1}} , \label{eq:Sexp}
\end{equation}
where $\delta_k$ denotes the dark-count rate at detector $k$.
Neglecting terms that include the product $\delta_1 \delta_2$, we
obtain the following approximation for the photon-induced
coincidence-count rate:
\begin{equation}
C^{(\mathrm{ph})} \approx \frac{C^{(\mathrm{raw})} R -
S_1^{(\mathrm{ph})} \delta_2 - \delta_1 S_2^{(\mathrm{ph})}}{R -
\delta_1 - \delta_2} , \label{eq:Cexp}
\end{equation}
where $C^{(\mathrm{raw})}$ is the raw coincidence-count rate.

\section{Experimental results}

\begin{figure}[t]
\centerline{\psfig{file=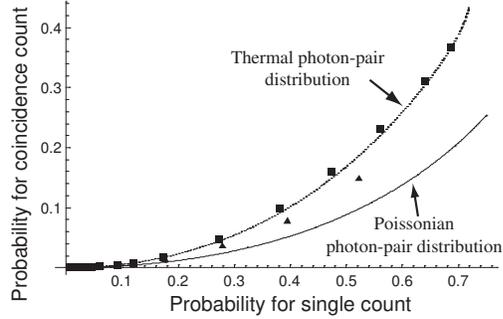,width=2.6in}}
\caption{Experimental data obtained with a 10-nm filter and pairs
of lenses with a focus length of 14.8~mm ($\blacksquare$) and
50~mm ($\blacktriangle$).} \label{fig:15and50mm}
\end{figure}

\begin{figure}[t]
\centerline{\psfig{file=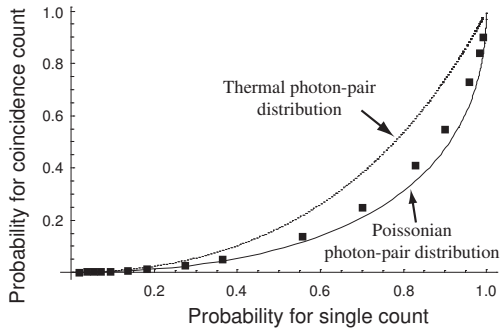,width=2.6in}}
\caption{Experimental results ($\blacksquare$) obtained with a
focus length of 50~mm and a 30-nm filter.} \label{fig:30nm}
\end{figure}

Using a laser emitting light with a wavelength of 1550~nm, the
transmittivity of the optical components was measured. Also the
detector efficiencies $\eta_1$ and $\eta_2$ were determined in
independent measurements. Plugging in these values into
Eqs.~(\ref{eq:Sk}) and (\ref{eq:C}), the single-count and
coincidence-count rates for a thermal or Poissonian photon-pair
distribution with a given average number of photon pairs could be
calculated. In Figs.~\mbox{\ref{fig:8mm}--\ref{fig:30nm}}, the
resulting relations between the probabilities for a single and a
coincidence count have been plotted as curves. Here, the
single-count probability is the probability for one or two
detectors to click, which is given by $(S_1 + S_2 - C)/R$.

In Fig.~\ref{fig:8mm}, the data obtained with a focal length of
8.2~mm and a 10-nm filter are presented. The different data points
correspond to different pump powers, and the effects of dark
counts have been compensated for. These measurements are seen to
be in agreement with a thermal photon-pair distribution. The
results obtained after changing the focal length to 14.8 and 50~mm
are shown in Fig.~\ref{fig:15and50mm}. We see that also with a
focal length of 14.8~mm, the data are in agreement with the
thermal distribution, whereas the longest focal length resulted in
a distribution lying between the thermal and Poissonian
photon-pair distributions. This can be understood considering the
dispersion between the pump and down-converted light in the
crystal. For a sufficiently large focus, i.e., for a sufficiently
long focal length of the lenses, photon pairs generated in
different locations within the focus will not emerge from the
crystal at the same time. The contributing processes can thus be
distinguished, resulting in a photon-pair distribution that is
closer to Poissonian.

In Fig.~\ref{fig:30nm}, we present the data obtained with a focal
length of 50~mm and a 30-nm filter. In this case, our measurements
are seen to be close to the Poissonian curve. This can be
explained by the fact that a broader filter allows for a shorter
coherence time of the transmitted down-converted
light,\cite{deRiedmatten,Mori} which makes it easier to
distinguish the generated photon pairs.

We note that a high number of photon pairs were generated in some
of our measurements. Assuming that the down-converted light
generated with a focal width of 14.8~mm indeed is described by a
thermal photon-pair distribution, and that the overall detection
efficiency equals the measured value of 2\%, the maximum average
number of generated photon pairs per pulse is found to be 60.
Similarly, the data presented in Fig.~\ref{fig:30nm} corresponds
to a maximum average number of photon pairs of 100, if the
generated photon pairs are assumed to have a Poissonian
distribution and the measured overall transmittivity of 2.8\% is
correct. In this case, the power of the down-converted light after
the filter was measured to be 2~nW using a simple power meter.

Finally, we note that, in contrast to our experiment with a
waveguide,\cite{Mori} we have not been able to fit the measured
single-count and coincidence-count rates as functions of pump
power to the theoretical expressions given in the previous
section. We believe that this is due to the high pump power and/or
the very high peak intensity achieved with the Titanium:sapphire
laser. This should increase the temperature of the crystal, and
consequently change the properties of the material and the
phase-matching conditions.

\section{Conclusions}

We have shown that even with only two single-photon detectors and
a low overall detection efficiency, insight into the photon
statistics can be gained. By changing the experimental conditions,
our experimental data could be varied from the expected results
for a thermal photon-pair distribution to those for a Poissonian
photon-pair distribution. This was explained by the increasing
degree of distinguishability within the distributed generation of
photon pairs.

\end{document}